\documentclass{appolb}
\usepackage{graphicx}
\usepackage{bm,amsmath,amssymb}

\newcommand \hmu {\hat{\mu}}
\newcommand \xb {\bar{x}}
\newcommand \cb {\bar{\chi}}

\newcommand{\pade}{Pad\'e}
\def\lsim{\raise0.3ex\hbox{$<$\kern-0.75em\raise-1.1ex\hbox{$\sim$}}}
\def\gsim{\raise0.3ex\hbox{$>$\kern-0.75em\raise-1.1ex\hbox{$\sim$}}}
\begin{document}
\title{Searching for the QCD critical point along the pseudo-critical/freeze-out line using {\pade}-resummed Taylor expansions of cumulants of conserved charge fluctuations%
\headtitle{Searching for the QCD critical point....}
\headauthor{Jishnu Goswami}
\thanks{Presented at Quark Matter 2022}%
}
\author{Jishnu Goswami \thanks{speaker}\address{RIKEN Center for Computational Science, 7-1-26 Minatojima-minami-machi, \\ Chuo-ku, Kobe, Hyogo 650-0047, Japan}\\ \vspace{4 mm} Frithjof Karsch, Christian Schmidt
\address{Fakult\"at f\"ur Physik, Universit\"at Bielefeld, D-33615 Bielefeld,
	Germany}
\\
\vspace{4 mm}
{Swagato Mukherjee}
\address{Physics Department, Brookhaven National Laboratory, Upton, NY 11973, USA}
}

\maketitle
\begin{abstract}
Using high-statistics datasets generated in (2+1)-flavor QCD calculations at finite 
temperature we construct estimators for
the radius of convergence from an eighth order series expansion of the pressure as well as the number density. 
We show that the estimator for pressure and number density will be identical in the asymptotic limit. In the vicinity of the pseudo-critical 
temperature, $T_{pc}\simeq 156.5$~MeV, we find the estimator of the radius of convergence to 
be $\mu_B/T \gsim\ 3$ for strangeness-neutral matter.
We also present results for the pole structure of the {\pade} approximants for the pressure 
at non-zero values of the baryon chemical potential and show that the 
pole structure of the [4,4]  {\pade} is consistent with not having a critical point at 
temperatures larger than $135~$MeV and a baryon chemical potential smaller than $\mu_B/T \sim \ 2.5$. 
\end{abstract}
  
\section{Introduction}
Taylor expansion and analytic continuation are the two most commonly used techniques
 to understand the properties of strongly interacting matter at non-zero values of the chemical potentials.
  Although both methods provide reliable estimates for thermodynamic observables at 
small chemical potentials, they suffer from systematic effects (truncation effects, 
limitation of analytic continuation ansatz etc.) at moderate to large chemical potentials as only 
few expansion coefficients are known \cite{review}. Hence, recently, there is a lot of effort 
going on in the lattice QCD community  aiming at an efficient resumation of the standard series expansions 
\cite{Gavai:2004sd,Mitra:2022vtf,Borsanyi:2022qlh,Dimopoulos:2021vrk} to get reliable estimates also at large chemical potentials. 
Here we will focus on the use of {\pade} approximations to resum the Taylor series to estimate the radius of 
convergence of Taylor series of pressure in (2+1)-flavor QCD at finite chemical potentials. A comparison of Taylor expansions and {\pade} resummation has been 
presented recently by 
the HotQCD collaboration in \cite{Bollweg:2022rps}.  

One of the central goals in QCD at large chemical potential is to find evidence 
for the existence of the so-called critical end point (CEP) in the QCD phase diagram. Phase transitions (critical points) are related to 
the singularities of the free energy on the real chemical potential axis, which one could estimate by analyzing the behavior of the expansion coefficients of Taylor series or by determining the poles of {\pade} approximants for thermodynamic observables obtained as 
derivatives of the partition function with 
respect to T or the chemical potentials \cite{PhysRev.87.404}. 
In the following we will elaborate on these ideas in the context of QCD at finite temperature and densities.  
Being forced to work with a finite number of Taylor coefficients {\pade} approximants are good choice as one can easily 
distinguish real and complex poles. Lattice QCD calculations at smaller-than-physical quark masses, 
combined with our model-based understanding of the QCD phase diagram, suggest that this critical point, 
if it exists, needs to be searched for at temperatures below/around the QCD chiral critical 
temperature($\sim 135~\rm{MeV}$) \cite{Halasz:1998qr,Karsch:2019mbv,HotQCD:2019xnw}. Thus we extend our calculations down to temperatures of
125 MeV and use the high statistics results for conserved charge cumulants up to 8th order, obtained 
by the HotQCD collaboration, to resum the Taylor expansions of the logarithm of the QCD partition function.

In the following sections, we will show that the poles one obtains from the diagonal [4,4] Pad\'e-approximants for 
8th order Taylor series of pressure in terms of baryonic chemical potential are complex at least for $T \gsim 140~$MeV, i.e. the singularity closest to the origin, which 
will control the radius of convergence of Taylor series, is in the complex plane. 
Of course, one has to confirm in the future, that this will be the case also for
higher order diagonal Pad\'e-approximants. 
This is consistent with the fact that the CEP does not exist for $T \gsim 140~$MeV.
\section{Taylor expansion and {\pade} approximants of isospin symmetric matter in (2+1)-flavor QCD}
The Taylor expansions for the 
pressure of $(2+1)$-flavor QCD is given by,
\begin{equation}
	\frac{P}{T^4} = \frac{1}{VT^3}\ln\mathcal{Z}(T,V,\vec{\mu}) = \sum_{i,j,k=0}^\infty%
\frac{\chi_{ijk}^{BQS}}{i!j!\,k!} \hmu_B^i \hmu_Q^j \hmu_S^k \; ,
\label{Pdefinition}
\end{equation}
with $\hat{\mu}_X\equiv \mu_X/T$.
Here, $\chi_{ijk}^{BQS}$ are 
derivatives of $P/T^4$ with respect to the corresponding chemical potentials,
$\vec{\mu}=(\mu_B, \mu_Q, \mu_S)$, evaluated at
$\vec{\mu}=\vec{0}$,
\begin{equation}
\chi_{ijk}^{BQS} =\left. 
\frac{1}{VT^3}\frac{\partial \ln\mathcal{Z}(T,V,\vec{\mu}) }{\partial\hmu_B^i \partial\hmu_Q^j \partial\hmu_S^k}\right|_{\vec{\mu}=0} \; ,
\; i+j+k\; {\rm even} \; .
\label{suscept}
\end{equation}

To study strangeness neutral ($n_S=0$) isospin-symmteric ($n_Q/n_B=0.5\Leftrightarrow \mu_Q=0$) matter,  we introduce
constraints on the strangeness chemical potentials, 
\begin{eqnarray}
        \hat{\mu}_S(T,\mu_B) &=& s_1(T)\hat{\mu}_B + s_3(T) \hat{\mu}_B^3+s_5(T) \hat{\mu}_B^5 +..  
\label{qs}
\end{eqnarray}
The expansion coefficients $s_i$ with $i=1,\ 3,\ 5, \ 7$ are given in \cite{HotQCD:2017qwq,Bazavov:2020bjn}.
Substituting $\mu_S$ by using Eq.~\ref{qs}
with $\mu_Q=0$ and using Eq.(\ref{Pdefinition}) we obtained, the Taylor series
for the $\hmu_B$-dependent part of the pressure and the net baryon-number density,
\begin{eqnarray}
\frac{P(T,\mu_B)-P(T,0)}{T^4}= \sum_{k=1}^{\infty} P_{2k}(T) \hmu_B^{2k} \ ; \frac{n_B(T,\mu_B)}{T^3} = \sum_{k=1}^{\infty} N_{2k-1}^B(T) \hmu_B^{2k-1}\;\; , \label{nXneutral}
\label{chineutral}
\end{eqnarray}

where, $N^B_{2k-1} = \frac{\bar{\chi}^{B,2k}_0}{(2k-1)!}$ and $\ P_{2k} = 2k N_{2k-1}=\frac{\bar{\chi}^{B,2k}_0}{2k!}$.

The simplest estimator, $r_{c,n}$, for the radius of convergence, 
$r_c=\lim\limits_{n\rightarrow \infty} r_{c,n}$, is obtained from the ratio of the subsequent, 
non-vanishing expansion coefficients. We define for pressure and number density respectively,
\begin{eqnarray}
r_{c,2k}^P=| P_{2k-2}/P_{2k}|^{1/2} &\text{and}&\ r_{c,2k}^{nB}=| N_{2k-3}/N_{2k-1}|^{1/2} \\
r_{c,2k}^P/r_{c,2k}^{nB} &=& \sqrt{[2k /(2k-2)]} = 1 + 1/k + O(k^2) \\
r_c&=&\lim_{k\to\infty}r^P_{c,2k}=\lim_{k\to\infty}r^{nB}_{c,2k}
\label{eq:pressure_number}
\end{eqnarray}

\begin{figure}
	\centering
	\includegraphics[width=5.8cm]{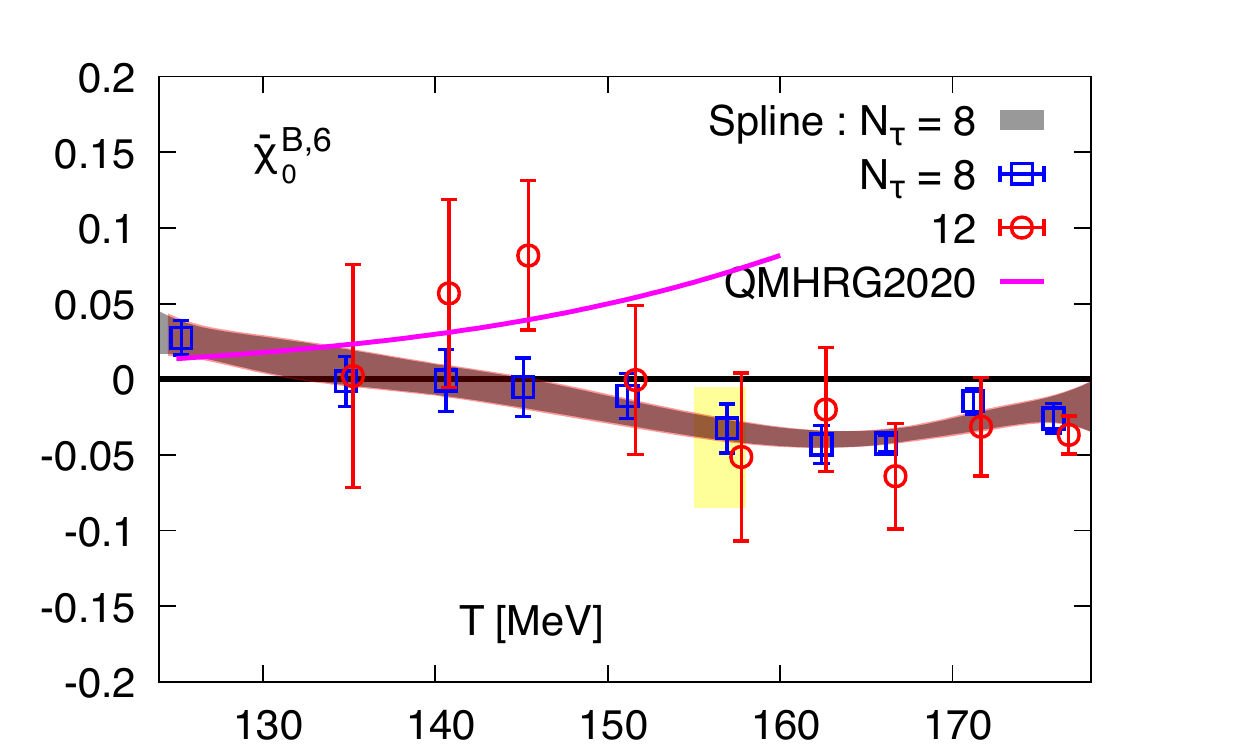}\hspace{-0.3cm}
	\includegraphics[width=5.8cm]{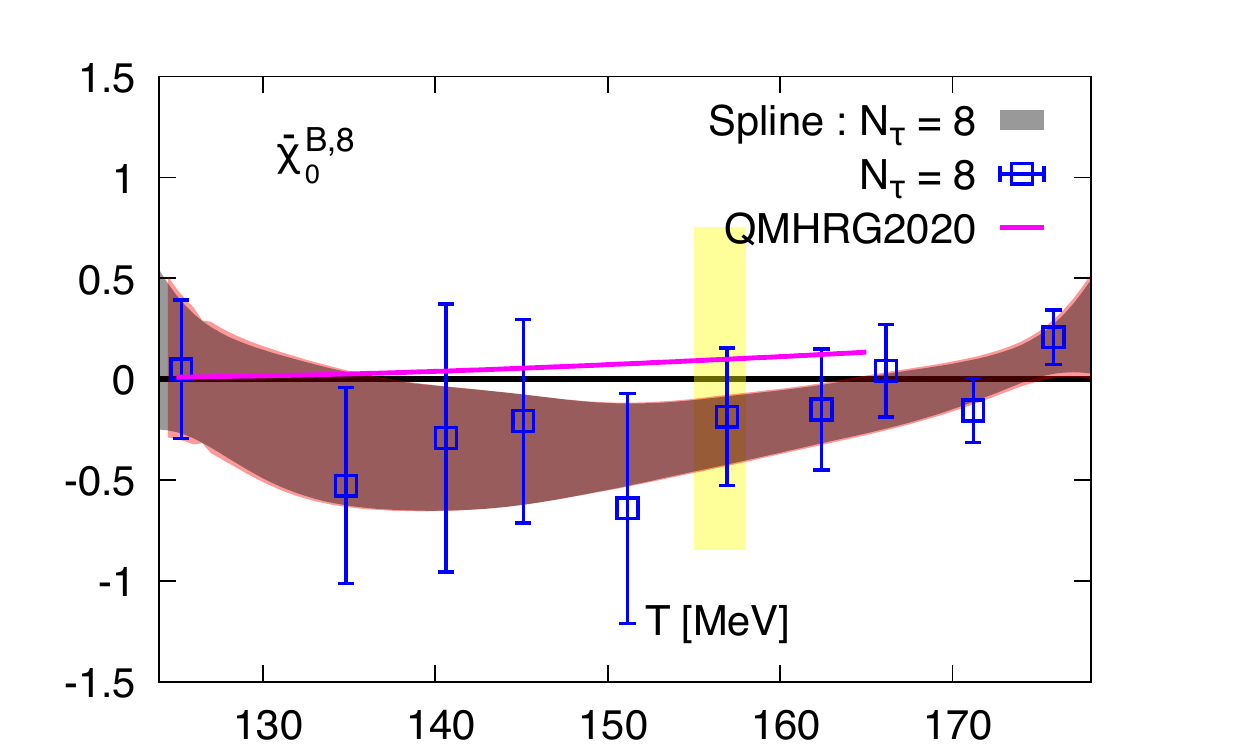}\hspace{-0.3cm}
	\caption{The sixth (left) and eighth (right) order cumulants,  contributing to the Taylor series of the pressure of
		(2+1)-flavor QCD  as function of $\hmu_B=\mu_B/T$ versus temperature. A Comparison with the hadron resonance gas model(QMHRG2020) \cite{Bollweg:2021vqf} is also shown.}
	\label{fig:suscept}
\end{figure}
In Fig. (\ref{fig:suscept}) we show the two highest order subsequent expansion coefficients. 
For these two expansion coefficients
we only used the spline interpolation of the datasets on $N_{\tau}=8$ lattice data for 
which we have about 1.5 million configurations for each temperature value.

\section{Searching for CEP using [4,4] Padé approximants}
Since the first two expansion coefficients in Eq. (\ref{nXneutral}) are strictly positive in the 
temperature range $T \in [135-175]~\rm{MeV}$, we rescale the expansion coefficients in the Taylor series, $c_{2k,2}=\frac{P_{2k}}{P_2} 
\left(\frac{P_2}{P_{4}}\right)^{k-1},~  \xb=\sqrt{\frac{P_{4}}{P_2}}$, to obtain,
\begin{eqnarray}
 \chi_0^B(T,\hmu_B)\frac{P_4}{P_2^2} &=&
\sum_{k=1}^{\infty} c_{2k,2} \xb^{2k}\; = \xb^2+\xb^4+ c_{6,2} \xb^6
+ c_{8,2} \xb^8
\;   \;
\end{eqnarray}
The $[2,2]$ and $[4,4]$ {\pade} can then be written as
\begin{eqnarray}
    \hspace*{-0.5cm}P[2,2] = \frac{\xb^2}{1-\xb^2}\;\; ,  \label{pd22} \ 
    P[4,4] =  \frac{(1-c_{6,2}) \xb^2 +
\left(1 - 2 c_{6,2} +c_{8,2} \right) \xb^4}{(1-c_{6,2})+
(c_{8,2}-c_{6,2}) \xb^2 + (c_{6,2}^2 - c_{8,2}) \xb^4} .
\label{pd44}
\end{eqnarray}

\begin{figure}
	\includegraphics[width=4.1cm]{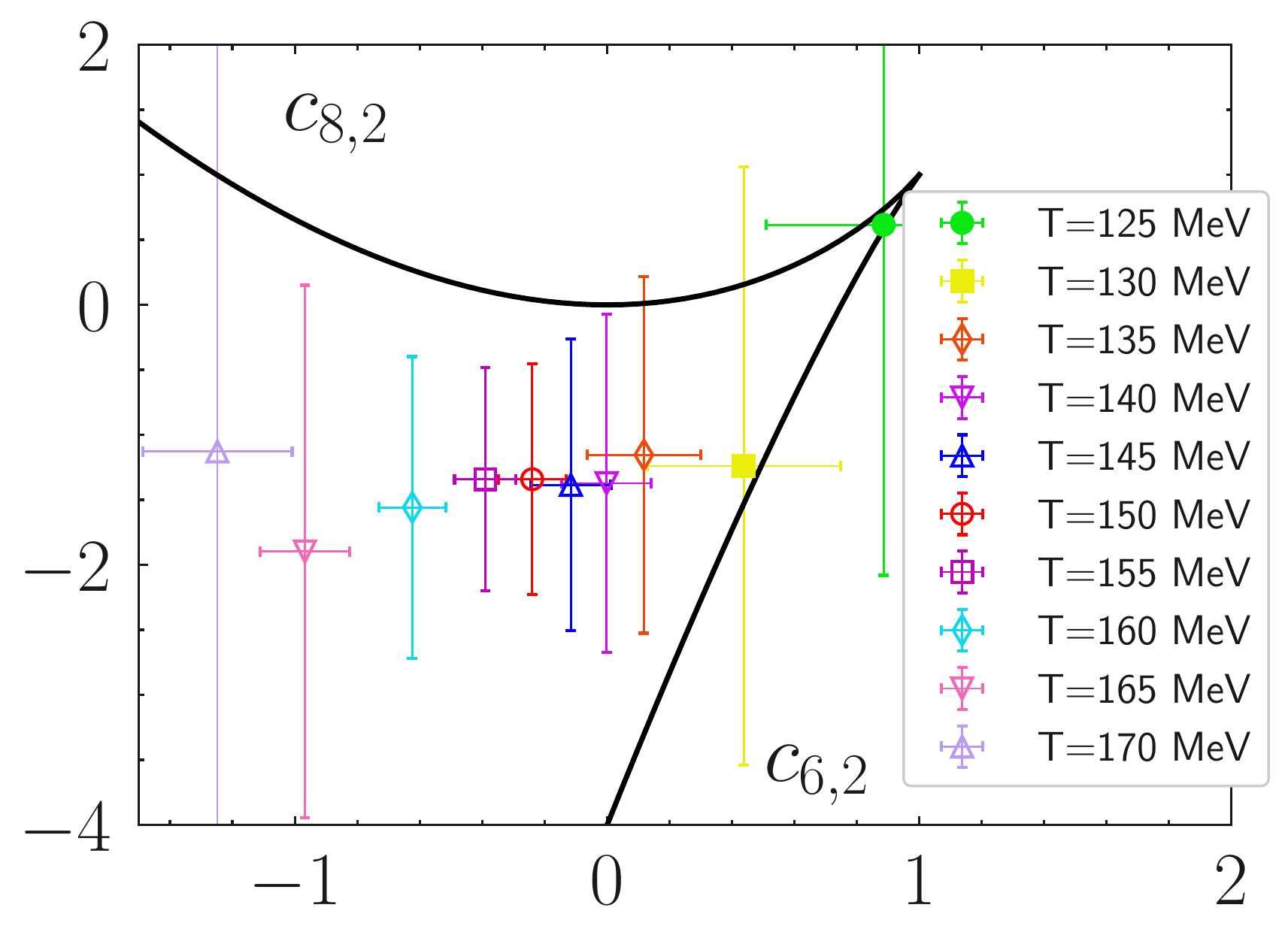}\hspace{-0.4cm}
	\includegraphics[width=5.2cm]{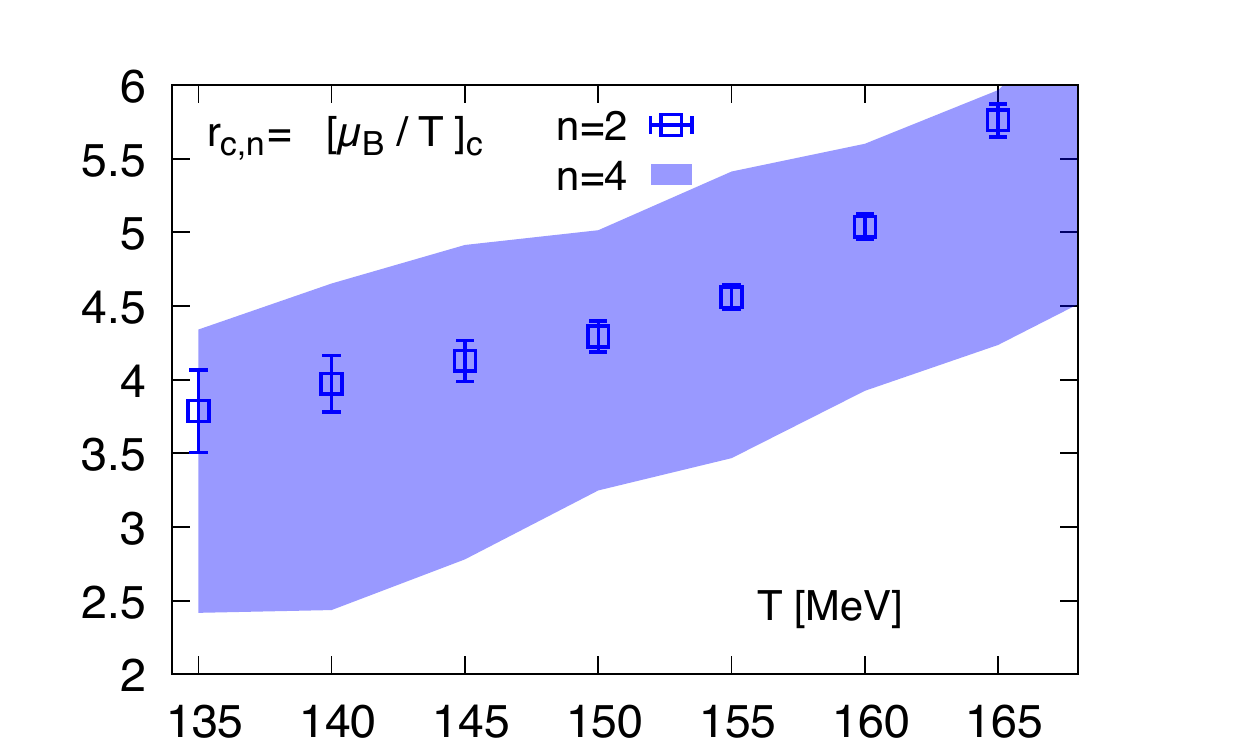}\hspace{-0.8 cm}
	\includegraphics[width=4.2cm]{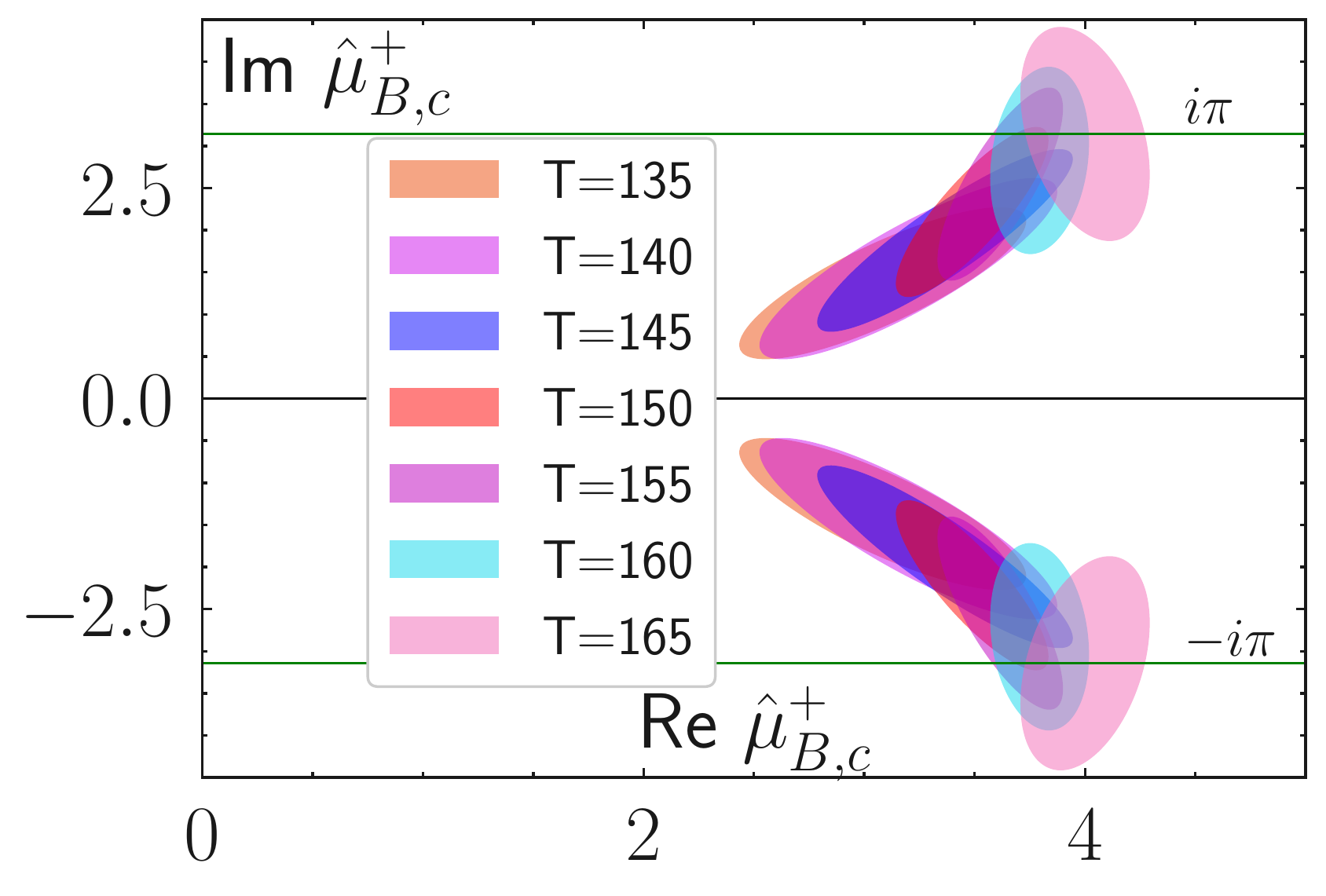} 
	\caption{$c_{8,2}$ vs $c_{6,2}$ on $N_\tau=8$ lattice
in the temperature range $125~{\rm MeV}< T < 175~{\rm MeV}$ (left), Magnitude of poles
nearest to the origin obtained from the [2,2] (squares) and [4,4] (bands) Pad\'e approximants (middle), Location of poles with ${\rm Re} (\mu_B)>0$ nearest to the origin obtained from the [4,4] Pad\'e approximants  in the complex $\hmu_B$-plane (right).}
	\label{fig:c8c6plane}
\end{figure}

The poles of the {\pade}s can be obtained by determining the roots of the denominators 
of Eq.(\ref{pd22}) as function of $\xb$. For, 
the case of the [2,2] {\pade} one gets $\xb^2=1$, {\it i.e.} for
$\mu_{B,c}\equiv r_{c,2}=\sqrt{12\cb^{B,2}_0/\cb^{B,4}_0}$, which 
is the standard ratio estimator for the radius of convergence. 
In the case of the [4,4] {\pade} there are four possibilities. Depending
on the values of $c_{8,2}$ and $c_{6,2}$ one will either find 
4 complex, or 2 real plus 2 imaginary, or 4 real, or 4 imaginary poles. 
Inside the triangular shaped regions bounded by black lines, shown in Fig.~\ref{fig:c8c6plane} (left), 
the poles are complex. We show in Fig.~\ref{fig:c8c6plane} (left), 
$c_{8,2}$ and $c_{6,2}$ obtained in (2+1)-flavor QCD. From that it can be 
established that one obtains 4-complex poles in the temperature range
$135~{\rm MeV}\le T\le 165~{\rm MeV}$. In Fig.~\ref{fig:suscept} our 
two highest expansion coefficients
$\cb_0^{B,6}$ and $\cb_0^{B,8}$ become positive at $T\simeq 125$~MeV, although 
errors still are large at lower temperatures. Hence, within 
our current statistical errors 
we cannot rule out a pair of real and/or purely imaginary 
poles at temperatures below $T=135$~MeV. The estimator for the 
radius of convergence for the complex poles can be written as,
\begin{eqnarray}
		r_{c,4} &=& \sqrt{\frac{12 \cb_0^{B,2}}{\cb_0^{B,4}}}  
		\left| 
		\frac{1-c_{6,2}}{c_{6,2}^2-c_{8,2}}\right|^{1/4}  \; ,\label{muBplane2} 
 \end{eqnarray}
which can be identify as the Mercer-Roberts estimator \cite{Mercer1990ACM} as long as the poles are complex. 
As seen from Fig.~(\ref{fig:c8c6plane}) 
(middle) estimates for the radius of convergence, obtained from diagonal {\pade} approximants lead to larger values with increasing temperature. We find
$\mu_B/T\gsim ~ [2.5-4]$ in the 
temperature range $T\sim[135-165]~\rm{MeV}$. We estimate  $\mu_B/T\gsim 3$ close to the pseudo-critical 
temperature $T_{pc}\sim 156.5 ~\rm{MeV}$. In Fig.~(\ref{fig:c8c6plane}) (right) 
we also show the location of  complex poles with a positive real part obtained in a 
temperature range between $T=[135:165]~\rm{MeV}$. They clearly show a tendency to move towards the real axis as the
temperature decreases. As mentioned earlier, our current statistical errors 
do not allow us to draw any conclusion about the nature of the poles at temperature 
 lower than $T\leq135~\rm{MeV}$.

\section{Conclusions}
Using diagonal {\pade} approximants of an eighth order 
Taylor series of (2+1)-flavor QCD
we  estimate a radius of convergence  of $\mu_B/T\gsim 3 $ 
close to the $T_{pc}$. We also show that the poles of the {\pade} approximants 
in the temperature range $T=[135:165]~\rm{MeV}$ are all complex, disfavoring the existence of a critical point at temperature larger than $T\sim135~\rm{MeV}$.
Furthermore, we also argued that for $T < 130~$MeV the [4,4]
Pad\'e-approximants can have real poles, which could 
 signal the occurrence of a phase transition, at lower temperatures. As the decrease of 
 the QCD 
 pseudo-critical temperature with increasing baryon 
 chemical potentials is small, such low temperatures 
 can only be reached for high baryon densities 
 $\mu_B/T\gsim 2.5$~MeV, i.e. at beam energies below the lowest bean energy used at RHIC in collider mode, $\sqrt{s}\lsim 7.7$~GeV.

\section*{Acknowledgements}
This work was supported by the Deutsche Forschungsgemeinschaft (DFG) - project number 315477589 - TRR 211, the European Union H2020-MSCA-ITN-2018-813942 (EuroPLEx), and the U.S. Department of Energy, Office of Science, Office of Nuclear Physics through i) the Contract No. DE-SC0012704 and (ii) the Office of Nuclear Physics and Office of Advanced Scientific Computing Research within the framework of Scientific Discovery through Advance Computing (SciDAC) award "Computing the Properties of Matter with Leadership Computing Resources''.


\end{document}